Biological Impact of Ozone Depletion at the End-Permian: A modeling study

Brian C. Thomas and Jacob M. Oberle

Washburn University, Department of Physics and Astronomy

1700 SW College Ave., Topeka, KS USA 66621

brian.thomas@washburn.edu

Abstract:

The end-Permian mass extinction is the most severe known from the fossil record. The most likely cause is massive volcanic activity associated with the formation of the Permo-Triassic Siberian flood basalts. A proposed mechanism for extinction due to this volcanic activity is depletion of stratospheric ozone, leading to increased penetration of biologically damaging Solar ultraviolet-B (UVB) radiation to Earth's surface. Previous work has modeled the atmospheric chemistry effects of volcanic emission at the end-Permian. Here we use those results as input for detailed radiative transfer simulations to investigate changes in surface-level Solar irradiance in the ultraviolet-B, ultraviolet-A and photosynthetically available (visible light) wave bands. We then evaluate the potential biological effects using biological weighting functions. In addition to changes in ozone column density we also include gaseous sulfur dioxide ($SO_2$) and sulfate aerosols. Ours is the first such study to include these factors and we find they have a significant impact on transmission of Solar radiation through the atmosphere. Inclusion of $SO_2$ and aerosols greatly reduces the transmission of radiation across the ultraviolet and visible wavelengths, with subsequent reduction in biological impacts by UVB. We conclude that claims of a UVB mechanism for this extinction are likely overstated.

Introduction

The end-Permian mass extinction is the most severe extinction known with at least 80% of marine, and 70% of terrestrial species disappearing from the fossil record[1–4]. The leading hypothesis for the cause of the "great dying" is volcanic activity associated with the formation of the Permo-Triassic Siberian flood basalts[5–8]. The mechanisms of extinction include climate change[6,9], ocean acidification[6], and ozone depletion[5–7,10,11]. Depletion of stratospheric ozone ($O_3$) leads to increased solar ultraviolet-B (UVB, 280-315 nm) irradiance at Earth's surface, which is known to have significant negative impacts on aquatic (both marine[12] and freshwater[13]) and terrestrial organisms[11,14–17]. Evidence of UVB-induced damage at the end-Permian comes from malformed spores and pollens[7,11,14,18] and recent experimental work has directly connected UVB exposure to malformation and sterilization of modern conifers[11]; a massive die-off of coniferous vegetation is associated with the extinction event[19].

Changes in atmospheric chemistry, including depletion of $O_3$ have been studied using a 2D chemistry-transport model[10] as well as a 3D Earth system model[5]. Both studies found significant $O_3$ depletion, but the level of depletion depends on details of the volcanic activity such as duration, type of eruption, and amounts and types of particular gasses emitted. In the most severe cases modeled, column density depletion of $O_3$ ranges from 60-85%, depending on latitude[5,10], while less severe cases yield depletion of 20-40%.

The gasses HCl, $CH_3Cl$, and $CH_4$ are particularly important in the ozone depletion process. The amount of these gasses emitted depends on the local geological conditions at the time and the nature of the eruptions. The ability to affect $O_3$ depends on whether these gasses can reach the stratosphere. Evidence indicates that the Siberian eruptions were likely extensive[20] and explosive[10] enough to inject significant amounts of these gasses into the stratosphere. In addition, the tropospheric lifetime of $CH_3Cl$ is long enough (greater than one year) for significant amounts to be transported into the stratosphere, even if it is not directly injected there by the eruptions. Strong evidence exists to support very large emission of volatiles, including the important $O_3$-depleting species, during the Siberian eruptions[7,8,20]. It has been shown that the high extinction rate during the period lines up best with sill intrusion through a volatile-rich basin[20]. Therefore, the combined geological evidence appears to make the most severe $O_3$ depletion cases[5,10] realistic.

Atmospheric modeling data

In this work we use results of the most recent and comprehensive modeling of atmospheric conditions at the end-Permian[5]. This modeling considered a variety of possible volcanic conditions, resulting in a range of $O_3$ depletion cases. We acquired data files for a case ("A2" in that work) that is slightly less severe than the most extreme model result, making our results somewhat conservative (from the perspective of $O_3$ depletion), given the possible range of impacts that are supported by the available geological evidence.

While $O_3$ is the most important atmospheric constituent controlling transmission of UVB (280-315 nm), volcanically produced $SO_2$ gas and sulfate ($H_2SO_4$) aerosols are also important. Broadly speaking, increased $SO_2$ column density and aerosol optical depth (AOD) lead to reduced transmission of UVB. A recent study[21] used the Tropospheric Ultraviolet and Visible (TUV) 1D radiative transfer model[22] (https://www2.acom.ucar.edu/modeling/tropospheric-ultraviolet-and-visible-tuv-radiation-model), to examine the effect on surface-level UVB irradiance by changes in these constituents. In this study, we use the methodology described in that work, applied to the specific conditions likely to have been in place during the Siberian Traps volcanic episode. While an earlier study[10] examined the effect of $O_3$ depletion alone at the end-Permian, ours is the first such study to include variations in $SO_2$ and sulfate aerosols, both of which are radiatively important and major atmospheric constituents during volcanic episodes.

Radiative Transfer Modeling

We use version 5.0 of the TUV model. TUV allows the user to modify a variety of parameter values, including date, time, location (latitude and longitude), altitude, and column density and vertical profile of $O_3$ and $SO_2$. Several aerosol optical parameters can be modified: total aerosol optical depth (AOD), single scattering albedo (SSA), Angstrom coefficient ($α$), a wavelength-independent scattering asymmetry factor ($g$), and vertical profiles. The default aerosol profile in TUV is for modern continental conditions, with AOD = 0.235, SSA = 0.99, $α$ = 1.0, and $g$ = 0.61.

We performed several simulations in order to evaluate the effects of varying different constituents. As a control, we use annual average $O_3$ column density values, as a function of latitude and longitude, from the control run in the modeling study by Black et al.[5] (hereafter "Black14") with globally uniform $SO_2$ column density of 0.05 DU and the default TUV aerosol parameters. The other simulation cases are presented in Table 1, where "Black14 A2" refers to (depleted) $O_3$ column density values from Siberian Traps volcanism case A2 in the Black14 study. In all cases the $SO_2$ and aerosol values are globally uniform. While the volcanic emission is not global, a recent study[23] found volcanic events that inject $SO_2$ into the stratosphere result in a roughly global distribution of gaseous $SO_2$ and sulfate aerosols, so our simplifying assumption is reasonable. In addition, our analysis will focus on broad comparisons rather than detailed geographic distribution of impacts.

We have chosen two "volcanic" aerosol cases. In both cases SSA = 0.99, $α$ = 2.0, $g$ = 0.74, and the altitude profile is a combined "flood" and "stratospheric" case; these parameter and profile choices are described in more detail in our previous work[21]. The combined low- and high-altitude profile is chosen as the best fit to the Siberian Traps eruption event[10,24]. In aerosol case 1 we use AOD = 0.5 and in case 2 AOD = 2.0;

these values are likely lower and upper limits for the period of active volcanism[25]. The other aerosol parameters are not varied since their effect on UVB transmission is relatively small, compared to that of AOD.[21] For $SO_2$ we take a value of 10.0 DU during the volcanic episode.[26]

The data we acquired from Black14 gives an annual average of $O_3$ column density as a function of latitude and longitude. In order to incorporate effects of seasonal variation on solar zenith angle we ran TUV at 1 day each month for 1 year, at local noon, at each longitude-latitude point. We then averaged over the 12 time points to give a single annual average value at each spatial point.

Results

We are interested in how changes in $O_3$ column density modeled for the end-Permian volcanic conditions affect surface-level Solar irradiance and subsequent biological impacts. In order to evaluate the impact of the reduced $O_3$, we have examined ratios of various values computed using the $O_3$ column density in the Black14 A2 volcanic case versus the Black14 control case.

Figure 1 shows the ratio of $O_3$ column density in the Black14 A2 volcanic case versus the Black14 control case. The depletion of $O_3$ is significant across the globe, with the largest change in the North polar region. In general, $O_3$ depletion is greatest in the polar regions due to factors including atmospheric transport and polar stratospheric clouds, and in this case the volcanic emissions are larger in the Northern hemisphere, leading to higher depletion in the North.

Figure 2 shows the ratio of computed surface-level UVB irradiance in the $O_3$-depleted (volcanic) case versus control. As expected, the increase in UVB irradiance tracks the decrease in $O_3$ column density. The increase is significant globally, with very large increases Northward of about 45° latitude, and Southward of about 55° latitude. Longitudinal variation is present in the $O_3$ variation due to geographical and volcanic emission variation in the climate model simulation. Since we are interested in more general impacts, the rest of our results and discussion will focus on zonally or globally averaged values.

Figure 3 shows the zonally averaged ratio of UVB irradiance for each of our volcanic cases versus control. The case labels refer to those listed in Table 1. For instance, the "$O_3$ + $SO_2$ + aer 2" case includes the volcanic case $O_3$ column density, our prescribed $SO_2$ column density, and our prescribed "aerosol 2" values. In every case, the same depleted (and control) $O_3$ column density distribution is used.

Notice in this plot that the latitude distribution of the ratios are all similar (as we would expect, since the $O_3$ column density distribution is the same for all cases), but the magnitude of the ratio is quite different between cases. Notably, the two cases that include our high optical depth "aerosol 2" settings show *reduced* irradiance compared to the control case, in the equatorial regions, where $O_3$ depletion is less severe. The high aerosol optical depth leads to much lower UVB irradiance compared to cases with lower optical depth.

In order to simplify comparisons between cases, Figures 4-12 show globally averaged ratios for several irradiance bands and biological weighting functions (BWFs). Note in Figure 4 that the high optical depth "aer2" cases show much lower ratios (as also seen in Figure 3). With $SO_2$ included with the high aerosol optical depth, the globally averaged ratio is 0.995, meaning the UVB irradiance is actually slightly *lower* than in the control case despite $O_3$ depletion.

Figure 5 shows results for UVA (315-400 nm), a biologically important waveband. In every case, except $O_3$ depletion only, there is less UVA irradiance compared to control, with less than half in the $SO_2$ plus high aerosol optical depth case. Similarly, Figure 6 shows results for photosynthetically available radiation (PAR, 400-700 nm), with lower irradiance compared to control in every case except $O_3$ depletion only.

Figure 7 shows results for erythema-weighted irradiance ratios. Erythema is a commonly used proxy for biological damage, but refers to skin damage and is therefore only relevant to relatively complex organisms. Erythema tracks closely with UVB, and the ratios here show a similar pattern, high for the $O_3$ depletion case, but much reduced under the addition of $SO_2$ and high optical depth aerosols.

Figures 8-10, show results for photosynthesis inhibition in phytoplankton. The largest increase is seen in Figure 8, using a BWF for inhibition of carbon fixation in a natural Antarctic phytoplankton community[27]. Figures 9 and 10 show results using BWFs for inhibition of photosynthesis in the phytoplankton species *Phaeo-dactylum* ("phaeo") and *Prorocentrum micans* ("proro"), respectively[28]. Again we note that inclusion of high optical depth aerosols leads to *reduced* inhibition of photosynthesis. However, with reduction in PAR (seen above) these results do not necessarily indicate an *increase* in productivity. Fully evaluating the effect on phytoplankton primary productivity will require more extensive modeling that incorporates a weighting function spanning the full UVB-UVA-PAR range[29,30].

A recent study[11] (hereafter "Benca18") has attempted to detect evidence of UVB exposure of gymnosperms at the end-Permian, using observations of malformed fossilized pollen grains compared to pollen grains of similar modern-day species that were exposed to specific levels of UVB irradiance under controlled conditions. That study used biologically weighted UVB exposure, calculated using a weighting function for land plants[31] (referred to here as "Caldwell71"). Figure 11 shows our results using this same BWF, again looking at globally averaged ratios of weighted irradiance in our various cases compared to the control. Notice that the vertical scale on this plot is larger than on the previous ones. This particular BWF is very sensitive to changes in UVB, much more so than another plant BWF[32] (referred to here as "Flint_Caldwell03") shown in Figure 12. Figure 13 shows the zonally averaged ratio for the Caldwell71 BWF; the pattern in latitude tracks that of the UVB ratio, but the range of values is much larger. However, as with other comparisons, the effect is greatly reduced with the addition of high optical depth aerosols and $SO_2$.

For their study of trees, Benca18 used daily integrated Caldwell71-weighted UVB flux. Their outdoor tree population (located in Berkeley, CA) was exposed to 7.2 kJ m$^{-2}$ day$^{-1}$. For Permian exposure, Benca18 used the only existing estimates of Permian plant damage-weighted UV levels, taken from a 2D atmospheric chemistry modeling study[10] (hereafter, "Beerling07"). In that work, modeled background (no volcanic perturbation) exposure was found to be 10-20 kJ m$^{-2}$ day$^{-1}$, and $O_3$ depleted values ranged 40-100 kJ m$^{-2}$ day$^{-1}$.

In order to make direct comparisons between our results and the Benca18 and Beerling07 values, we ran TUV hourly, one day per month, for one year, using zonally averaged $O_3$ column density values. The hourly Caldwell71-weighted irradiance results were then daily-integrated. Our Permian control case gives maximum flux values of around 8 kJ m$^{-2}$ day$^{-1}$, which is similar to the Benca18 modern day outdoor exposure. However, our value here is smaller than the Permian control values from Beerling07.

For our Permian $O_3$-depleted only case, we found maximum flux values of around 20 kJ m$^{-2}$ day$^{-1}$, which is significantly smaller than the Beerling07 maximum value. For our Permian $O_3$-depleted with $SO_2$ and high optical depth aerosol case, we found maximum flux values of around 6 kJ m$^{-2}$ day$^{-1}$, smaller even than the control case.

Overall then, even considering only $O_3$ depletion, we find lower Caldwell71-weighted flux values compared to Beerling07, and much smaller values when including $SO_2$ and high optical depth aerosols.

Discussion and Conclusions

Our study represents the most detailed examination to date of surface-level exposure to Solar UVB under probable end-Permian conditions. We have used a state-of-the-art radiative transfer model with best-estimate conditions for this time period. Broadly, we find that there is a large increase in UVB exposure, and subsequent biological impacts, in the $O_3$ depleted case – as much as an order of magnitude higher in North polar areas where depletion is greatest.

However, we also find that inclusion of $SO_2$ gas and sulfate aerosols has a major effect, greatly reducing exposure, in some cases even below what would be expected in an normal $O_3$ column density case. This is a significant result because it indicates that UVB exposure may not be as much of a contributor to the end-Permian extinction event as has been previously claimed.

We also find that previous plant damage weighted flux values given in the literature for this time period may be over estimated. The cause for the difference between our results and those of Beerling07 is not obvious and difficult to track down since few details are given in that work as to how flux values were arrived at. However, the most likely cause is simplification or neglect of processes such as scattering by that work, where are accurately included in our modeling. A study[33] comparing simplified radiative transfer methods to results from the TUV model found that the simplified method overestimated UVB irradiance by a factor of about 2, depending on specific parameter values chosen.

Our results indicate that observations of malformed fossilized pollen[11] may not actually track increased UVB exposure, but may instead show exposure to heightened surface-level $SO_2$ concentration[10], which also certainly accompanies major volcanic activity.

There are many potential sources of uncertainty in this work, most importantly a lack of knowledge about the $SO_2$ and aerosol conditions during this period. Our results likely bracket the maximum and minimum values that are realistic for the end-Permian volcanic episodes.

Finally, we note that ours is the first study to examine changes in surface level irradiance in the UVA and PAR bands for the end-Permian, and we find reduced levels in all cases except the $O_3$ depleted only case. The biological implications of this are not obvious and will require more extensive modeling that includes effects of the full UVB-UVA-PAR wavelength range.

Acknowledgements:

The authors thank Benjamin Black for providing data and assistance in processing that data into a format usable for this study.

Analysis and plotting was done using the NCAR Command Language (Version 6.5.0) [Software]. (2018). Boulder, Colorado: UCAR/NCAR/CISL/TDD. http://dx.doi.org/10.5065/D6WD3XH5

(Iceland). *J. Geophys. Res. Atmos.* **120**, 9739–9757 (2015).

27. Boucher, N. *et al.* Icecolors '93: Biological weighting function for the ultraviolet inhibition of carbon fixation in a natural antarctic phytoplankton community. *Antarct. J. Rev.* 272–275 (1994).

28. Cullen, J. J., Neale, P. J. & Lesser, M. P. Biological Weighting Function for the Inhibition of Phytoplankton Photosynthesis by Ultraviolet Radiation. *Science (80-. ).* **258**, 646–650 (1992).

29. Neale, P. J. & Thomas, B. C. Solar irradiance changes and phytoplankton productivity in earth's ocean following astrophysical ionizing radiation events. *Astrobiology* **16**, (2016).

30. Neale, P. J. & Thomas, B. C. Inhibition by ultraviolet and photosynthetically available radiation lowers model estimates of depth-integrated picophytoplankton photosynthesis: global predictions for Prochlorococcus and Synechococcus. *Glob. Chang. Biol.* **23**, 293–306 (2017).

31. Caldwell, M. M. SOLAR UV IRRADIATION AND THE GROWTH AND DEVELOPMENT OF HIGHER PLANTS. in *Photophysiology* 131–177 (Elsevier, 1971). doi:10.1016/b978-0-12-282606-1.50010-6

32. Flint, S. D. & Caldwell, M. M. A biological spectral weighting function for ozone depletion research with higher plants. *Physiol. Plant.* **117**, 137–144 (2003).

33. Thomas, B. C., Neale, P. J. & Snyder, B. R. Solar Irradiance Changes and Photobiological Effects at Earth's Surface Following Astrophysical Ionizing Radiation Events. *Astrobiology* **15**, 207–220 (2015).
Thomas & Oberle; Ozone Depletion at the End-Permian                                    Page 18 of 32

Table 1: Atmospheric constituent cases. "volc" refers to our volcanic aerosol and $SO_2$ profiles; "aer 1" and "aer 2" refer to our two aerosol cases.

| Case | $O_3$ | $SO_2$ (DU) | $SO_2$ profile | AOD | SSA | $\alpha$ | g | aerosol profile |
|---|---|---|---|---|---|---|---|---|
| Control | Black14 Control | 0.05 | TUV default | 0.235 | 0.99 | 1.0 | 0.61 | TUV default |
| $O_3$ only | Black14 A2 | 0.05 | TUV default | 0.235 | 0.99 | 1.0 | 0.61 | TUV default |
| $O_3$ + aer 1 | Black14 A2 | 0.05 | TUV default | 0.5 | 0.99 | 2.0 | 0.74 | volc |
| $O_3$ + aer 2 | Black14 A2 | 0.05 | TUV default | 2.0 | 0.99 | 2.0 | 0.74 | volc |
| $O_3$ + $SO_2$ | Black14 A2 | 10.0 | volc | 0.235 | 0.99 | 1.0 | 0.61 | TUV default |
| $O_3$ + $SO_2$ + aer 1 | Black14 A2 | 10.0 | volc | 0.5 | 0.99 | 2.0 | 0.74 | volc |
| $O_3$ + $SO_2$ + aer 2 | Black14 A2 | 10.0 | volc | 2.0 | 0.99 | 2.0 | 0.74 | volc |

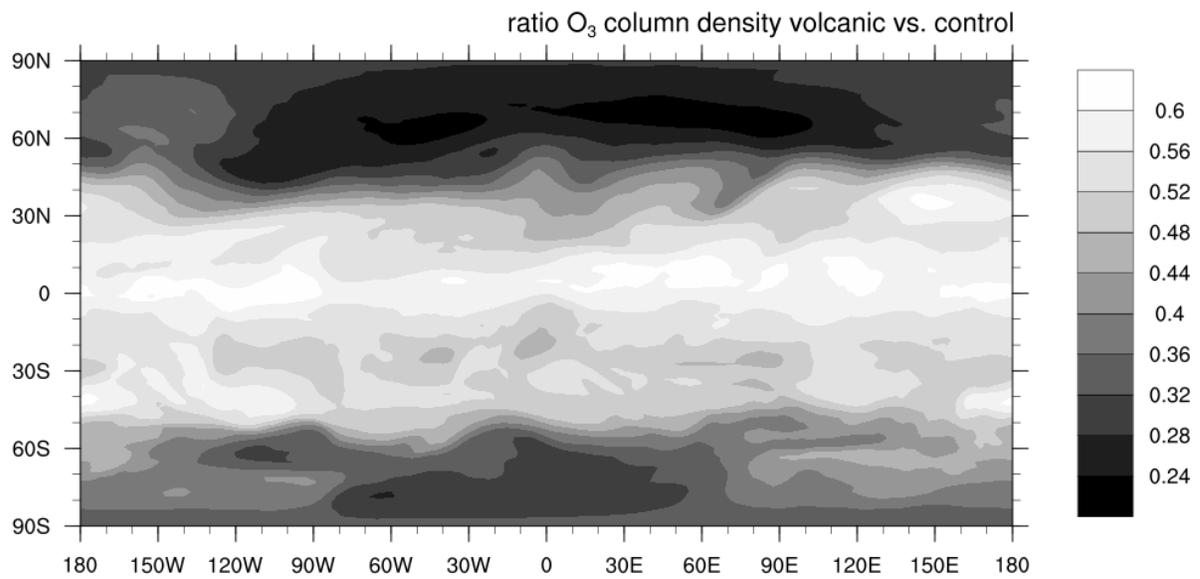

Figure 1 – Ratio of $O_3$ column density in the Black14 A2 volcanic case versus the Black14 control case.

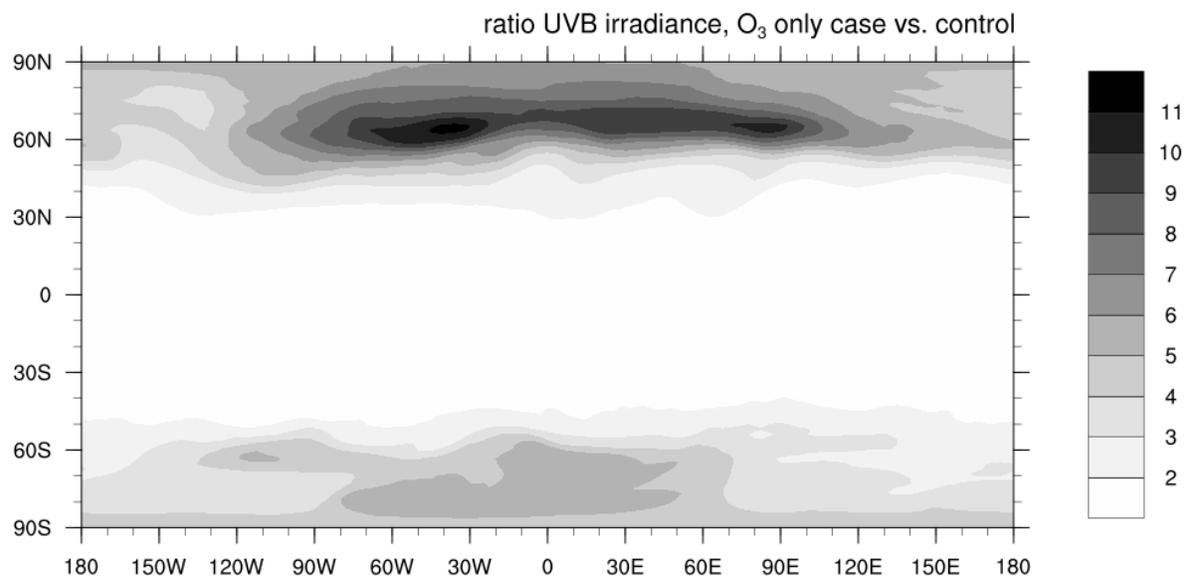

Figure 2 – Ratio of computed surface-level UVB irradiance in the $O_3$-depleted (volcanic) case versus control.

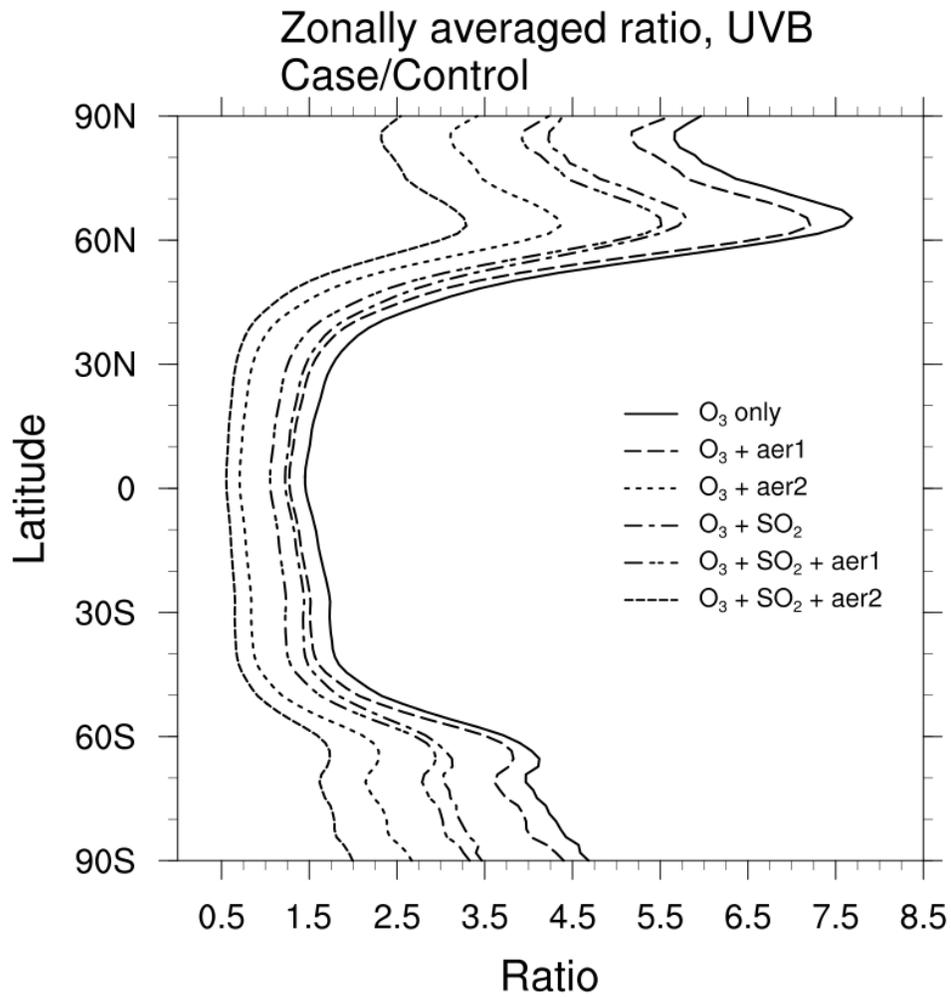

Figure 3 - Zonally averaged ratio of UVB irradiance for each of the volcanic cases (see Table 1) versus control.

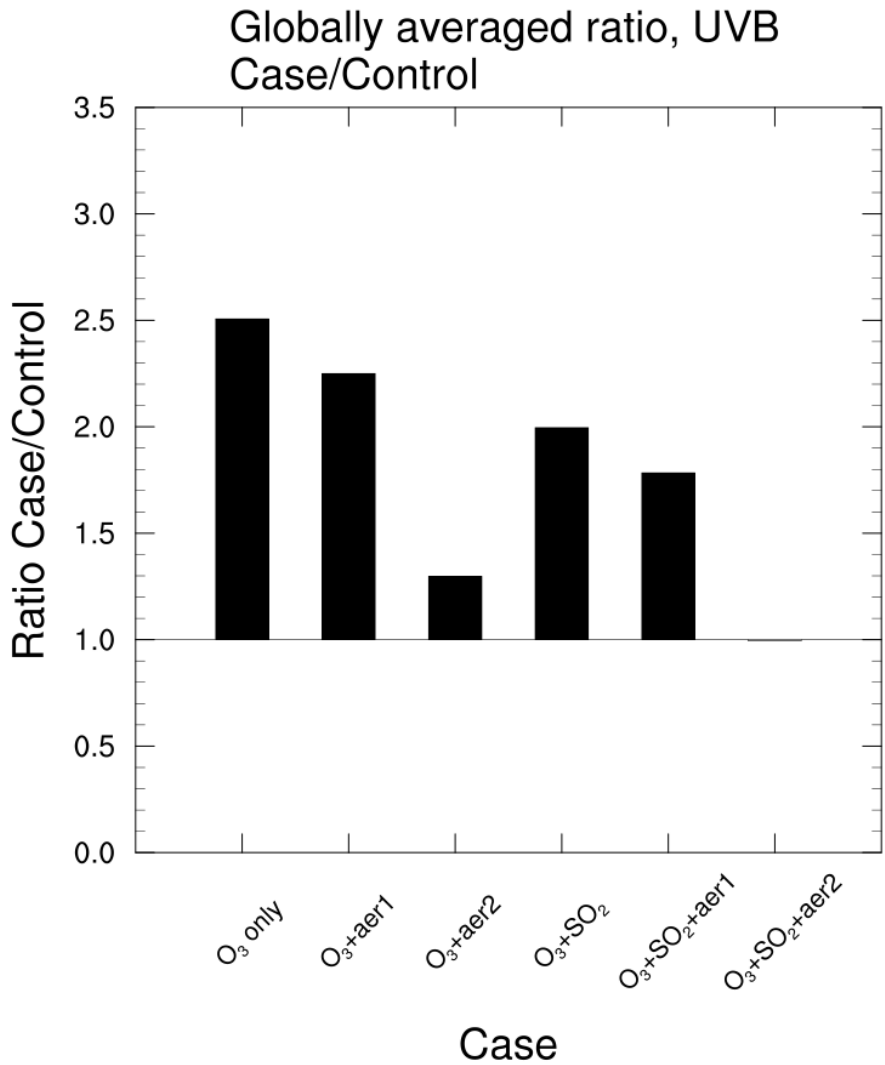

Figure 4 – Globally averaged ratio of UVB irradiance for each of the volcanic cases (see Table 1) versus control.

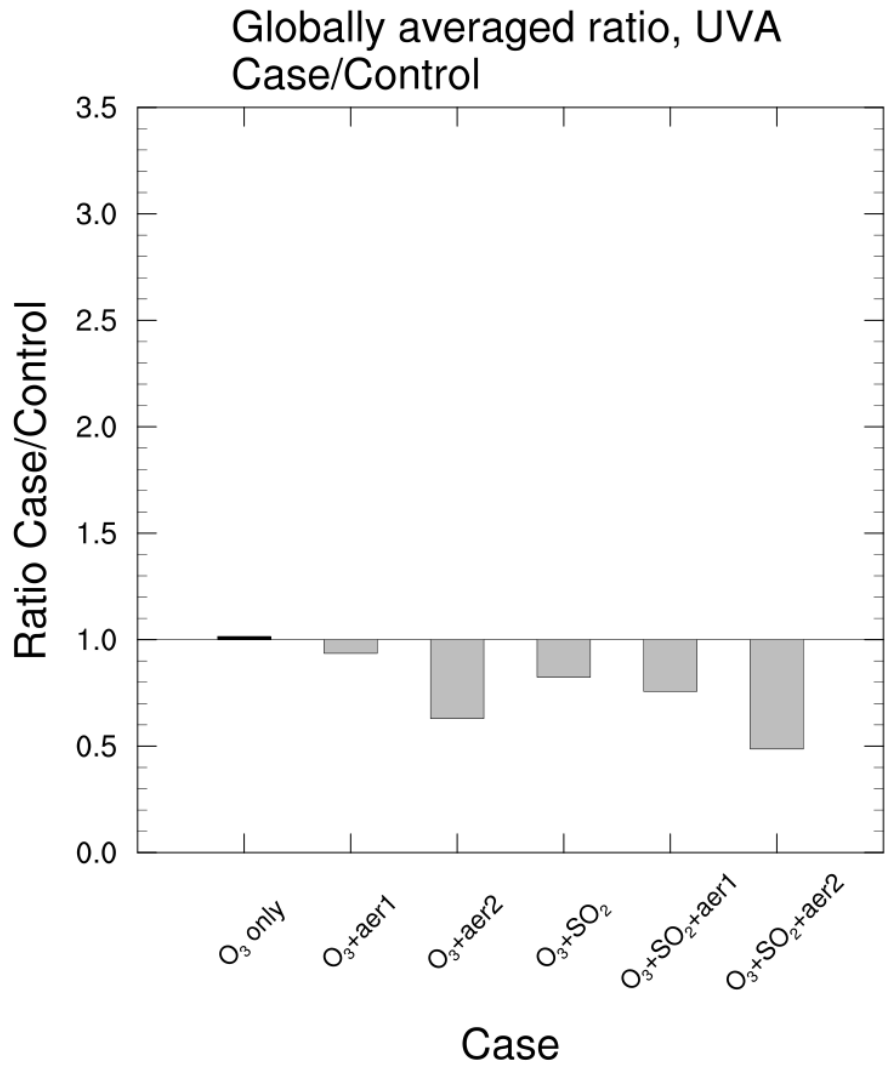

Figure 5 – Globally averaged ratio of UVA irradiance for each of the volcanic cases (see Table 1) versus control.

Thomas & Oberle; Ozone Depletion at the End-Permian					Page 24 of 32

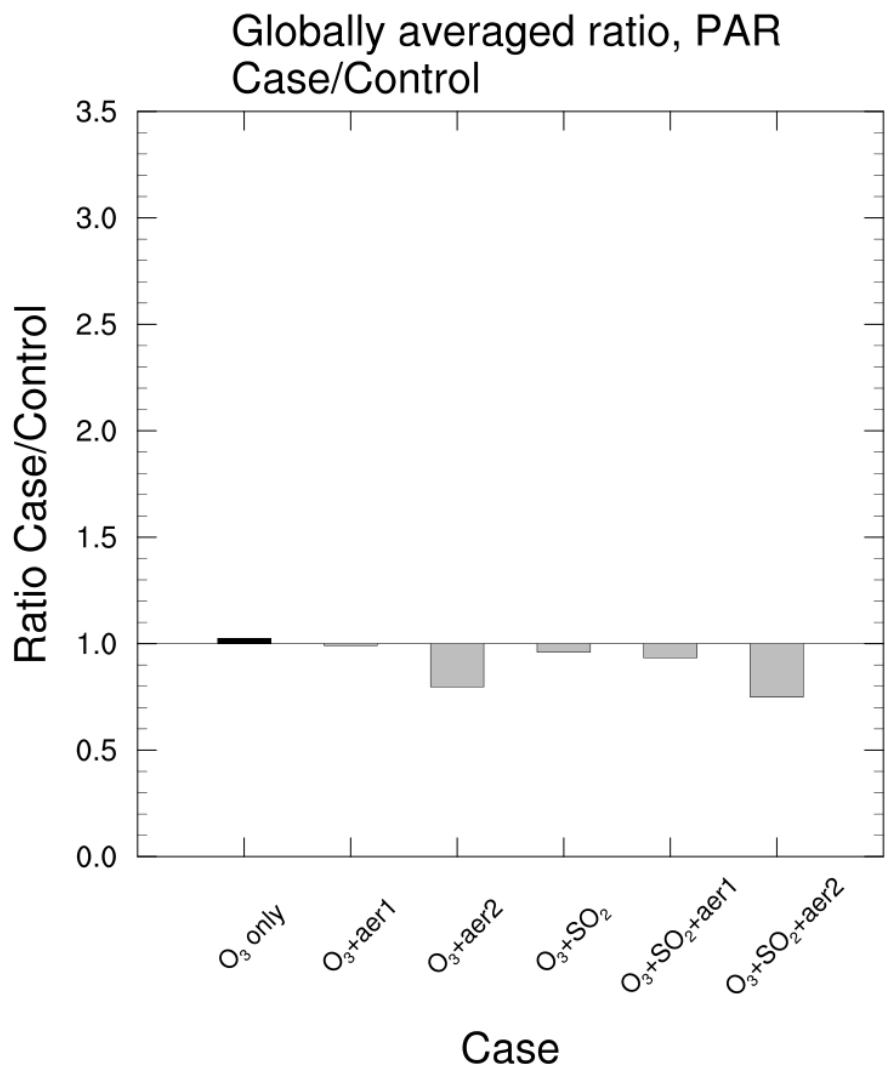

Figure 6 – Globally averaged ratio of PAR irradiance for each of the volcanic cases (see Table 1) versus control.

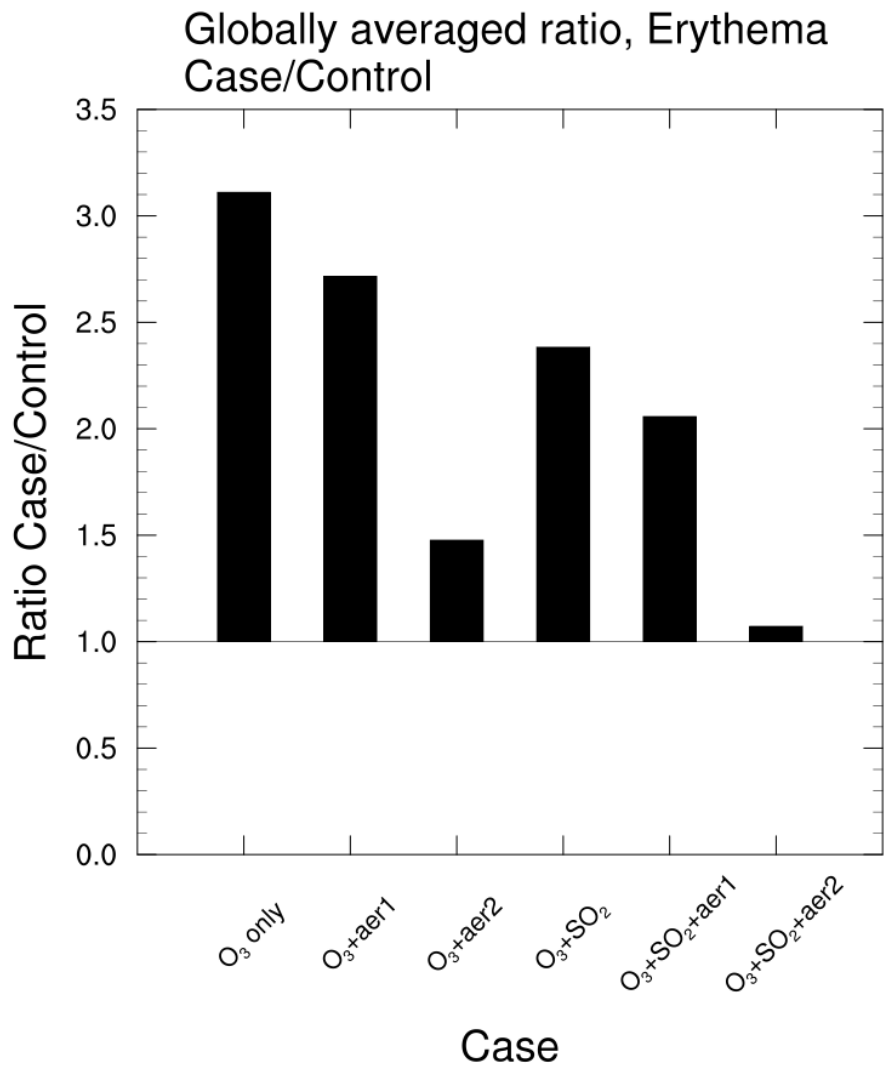

Figure 7 – Globally averaged ratio of Erythema-weighted irradiance for each of the volcanic cases (see Table 1) versus control.

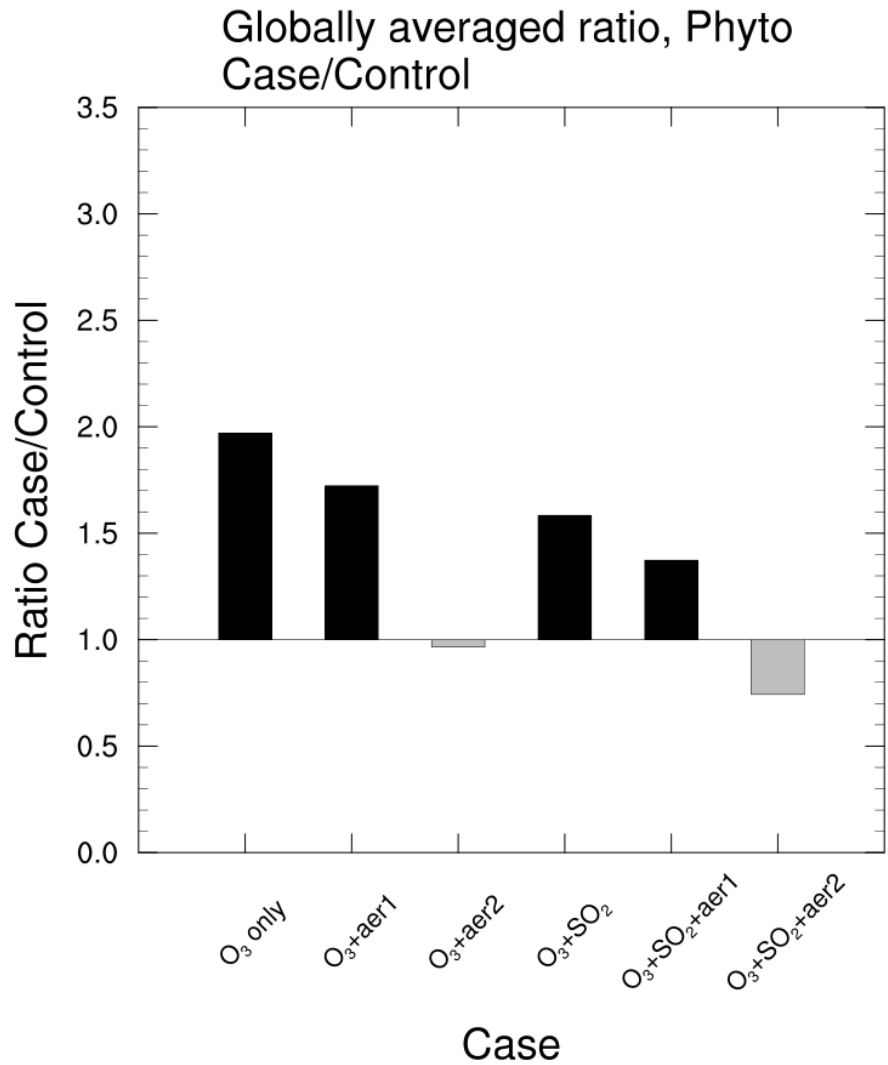

Figure 8 – Globally averaged ratio of phytoplankton photosynthesis inhibition-weighted irradiance (using a BWF for inhibition of carbon fixation in a natural Antarctic phytoplankton community[27]) for each of the volcanic cases (see Table 1) versus control.

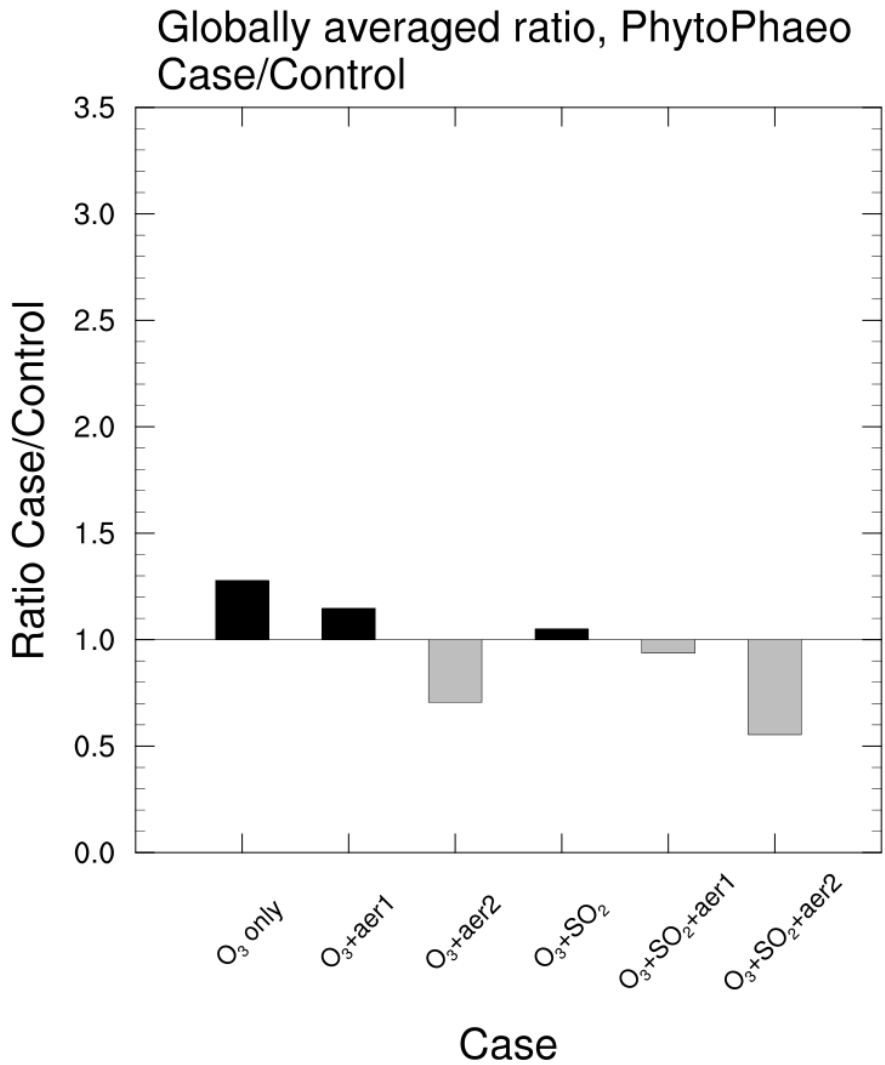

Figure 9 – Globally averaged ratio of phytoplankton photosynthesis inhibition-weighted irradiance (using a BWF for *Phaeo-dactylum*[28]) for each of the volcanic cases (see Table 1) versus control.

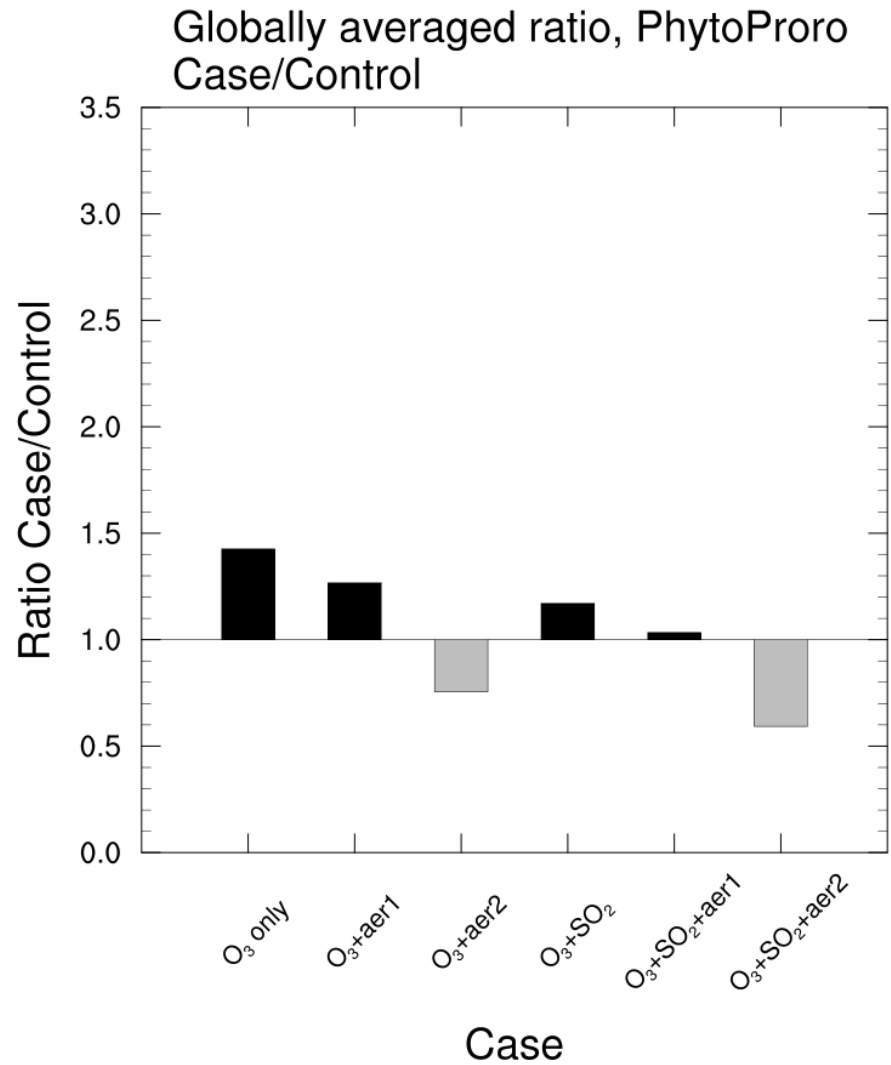

Figure 10 – Globally averaged ratio of phytoplankton photosynthesis inhibition-weighted irradiance (using a BWF for *Prorocentrum micans* [28]) for each of the volcanic cases (see Table 1) versus control.

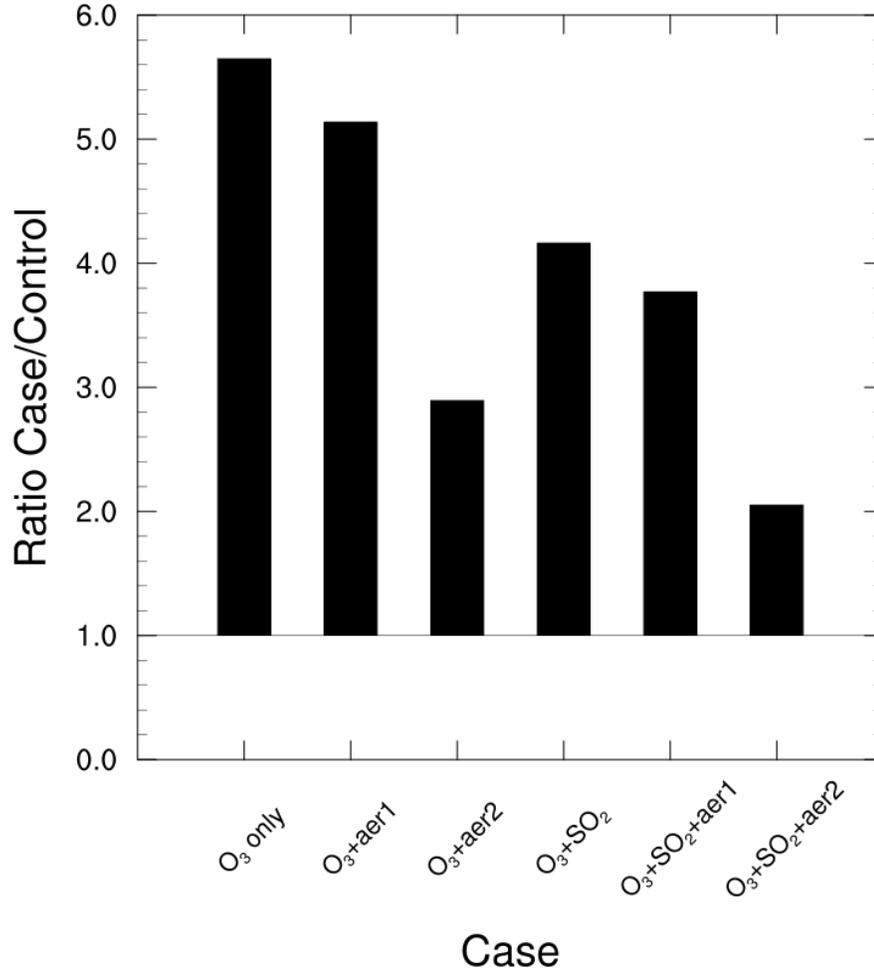

Figure 11 – Globally averaged ratio of plant damage-weighted[31] irradiance for each of the volcanic cases (see Table 1) versus control. Note the vertical axis scale here is larger than in previous similar plots.

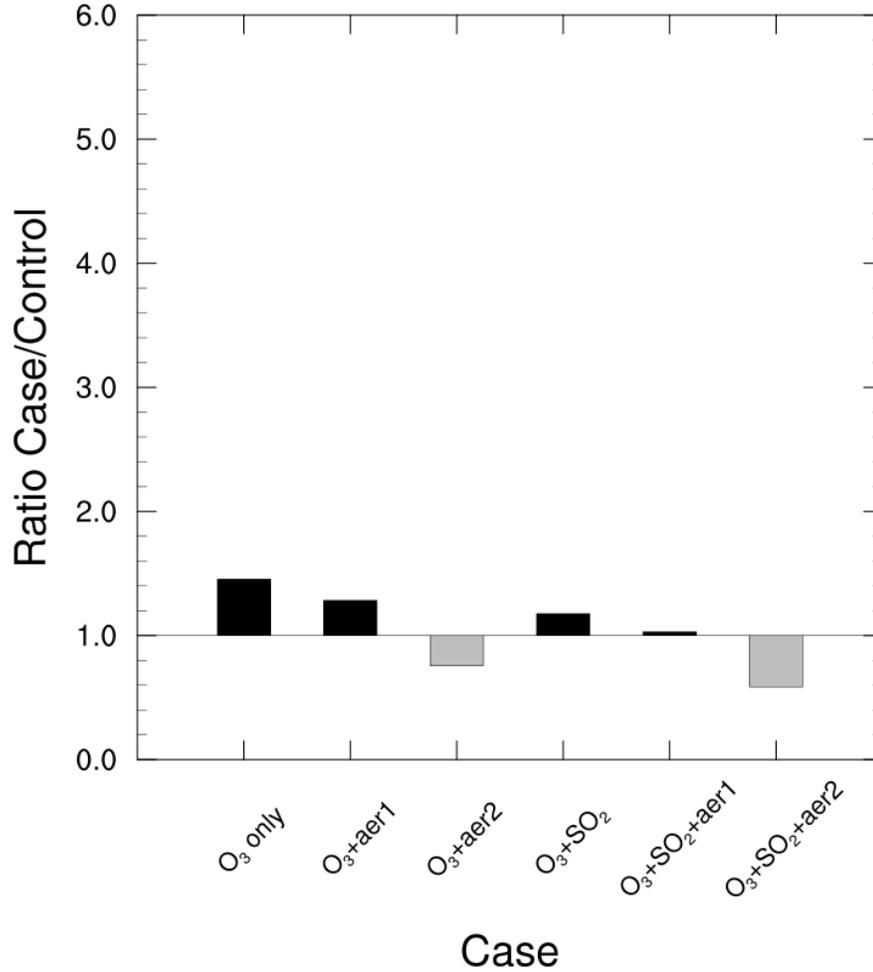

Figure 12 – Globally averaged ratio of plant damage-weighted[32] irradiance for each of the volcanic cases (see Table 1) versus control. Note the vertical axis scale here is chosen for comparison with Figure 11.

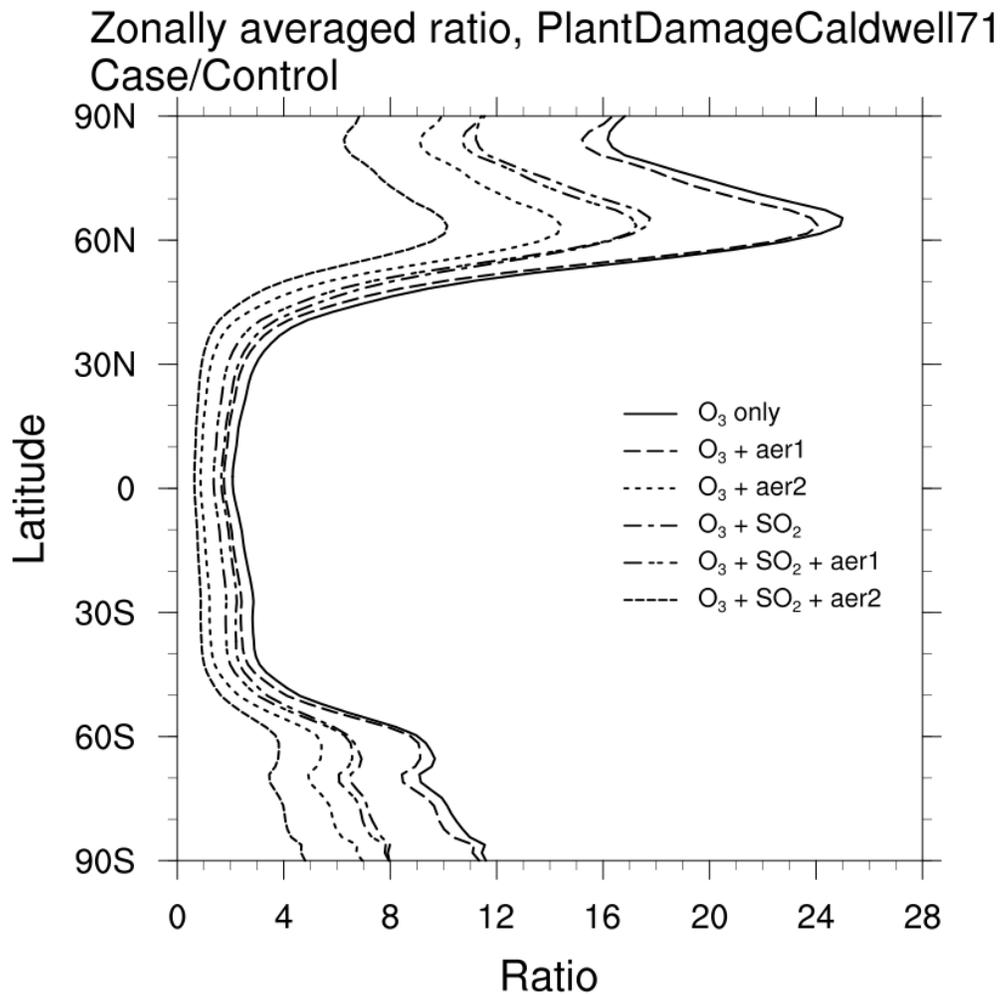

Figure 13 - Zonally averaged ratio of plant damage-weighted[31] irradiance for each of the volcanic cases (see Table 1) versus control. Note the horizontal axis scale is larger than the comparable Figure 3.